\documentclass[11pt]{article}
\usepackage[dvips]{graphicx}
\usepackage{url}
\usepackage{hyperref}

\usepackage{amsmath}
\usepackage{t1enc}
\usepackage[utf8]{inputenc}
\usepackage[toc,page]{appendix}
\usepackage{caption}
\usepackage{amsfonts,amssymb,xfrac,nicefrac,esint,float}
\usepackage{xcolor} 
\usepackage{csquotes}

\newcommand\nc{\newcommand}
\nc{\todo}[1]{}
\renewcommand{\todo}[1]{\textcolor{blue}{\texttt{[:~#1~:]}}}
\nc{\todoo}[1]{}
\renewcommand{\todoo}[1]{\textcolor{red}{\texttt{[:~#1~:]}}}
\nc{\SzM}[1]{\todoo{#1 -- \emph{SzM}}}
\nc{\VP}[1]{\todo{#1 -- \emph{VP}}}

\nc\lat\textit
\nc\ie{\lat{i.e.,\ }} \nc\etal{\lat{et al.\ }} \nc\etc{\lat{etc.\ }}
\nc\eg{\lat{e.g.,\ }} \nc\insitu{\lat{in situ}} \nc\QED{\lat{Q.E.D.}}
\nc\cf{cf.\ } \nc\wrt{w.r.t.\ }
\nc\adden{extra energy}
\nc\difficoeff{A}

\def\re#1{(\ref{#1})} 

\nc\ttensor\mathbf
\nc\Tensor\boldsymbol
\nc\ident{\mathbf1}
\nc\zero{\ttensor{0}}

\nc\dd{{\rm d}}
\nc\pd{\partial}
\nc\pdt[1]{\frac{\pd #1}{\pd t}}
\nc\pder[2]{\frac{\pd #1}{\pd #2}}
\nc\pderr[3]{\left. \frac{\pd #1}{\pd #2} \right|_{#3}}
\nc\ppder[2]{\frac{\pd #1}{\pd \left( #2 \right) }}
\nc\ppderr[3]{\left. \frac{\pd #1}{\pd \left( #2 \right) } \right|_{#3}}
\nc\qdot[1]{\left( #1 \right) \dot{\vphantom{\left( #1 \right)}}}

\nc\qrho{\varrho}
\nc\qphi{\varphi}

\nc\qJE{J_E}
\nc\qqJE{\ttensor{J}_E}
\nc\qJU{J_U}
\nc\qqJU{\ttensor{J}_U}
\nc\qJS{J_S}
\nc\qqJS{\ttensor{J}_S}

\nc\he{\varepsilon}
\nc\qqh{{\ttensor{h}}}
\nc\qP{{\mathsf{P}}}
\nc\qqP{{\ttensor{P}}}
\nc\qqPi{{\Tensor{\Pi}}}
\nc\qqq{{\ttensor{q}}}
\nc\qqr{{\ttensor{r}}}
\nc\qv{{\mathsf{v}}}
\nc\qqv{{\ttensor{v}}}
\nc\qm{{\mathsf{m}}}
\nc\qM{{\mathsf{M}}}

\nc\bbS{\mathbb{S}}
\nc\cB{\mathcal{B}}

\nc\be{\Gamma_{\hskip -0.5ex e}}
\nc\br{\Gamma_{\hskip -0.5ex\qrho}}
\nc\bgr[1]{\Gamma_{\hskip -0.5ex \nabla \qrho}^{#1}}
\nc\bbgr{\Tensor{\Gamma}_{\hskip -0.5ex \nabla \qrho}}
\nc\bp{\Gamma_{\hskip -0.5ex \qphi}}
\nc\bgp[1]{\Gamma_{\hskip -0.5ex \nabla \qphi}^{#1}}
\nc\bbgp{\Tensor{\Gamma}_{\hskip -0.5ex \nabla \qphi}}
\nc\bv[1]{{{\Gamma_{\hskip -0.5ex{\qqv}}}^{}}^{}_{#1}}
\nc\bbv{{{\Tensor{\Gamma}_{\hskip -0.5ex{\qqv}}}^{}}^{}}

\nc\qf{_{\rm fl}}
\nc\id{_{\rm per}}

\nc\ee{{\rm e}}
\nc\ii{{\rm i}}

\nc\qS{\mathsf{S}}

\title{\fontsize{14}{12}{\bf SOME REMARKS ON THE OBJECTIVITY AND THERMODYNAMIC CONSISTENCY OF KORTEWEG-TYPE FLUIDS}\footnote{This paper is dedicated to Liliana Restuccia, because she is already closer to 60 than 50.}}

\author{P. V\'an$^{1,2}$ \footnote{
		{\tt van.peter@wigner.hu} $^1$Department of Theoretical Physics, HUN-REN Wigner Research Centre for Physics, H-1525 Budapest, Konkoly Thege Miklós u. 29-33., Hungary; 
		and  $^2$Department of Energy Engineering, Faculty of Mechanical Engineering,  Budapest University of Technology and Economics, H-1111 Budapest, Műegyetem rkp. 3., Hungary} }
\date{~}

\usepackage[twoside]{fancyhdr}
\begin{document}

\maketitle
	
\bigskip

\begin{abstract}	
	In this note we compare the entropy principle and the objectivity arguments in the methodologies of Dunn and Serrin \cite{DunSer85a} and in the more recent weakly nonlocal thermodynamic analysis of Korteweg-type fluids in \cite{Van23a}. It is concluded that  the different objectivity approaches lead to the same constitutive functions, and that the difference in the thermodynamically compatible pressure tensors of perfect Korteweg fluids is due to different symmetry requirements.
\end{abstract}


\section{Introduction}
	
Material frame indifference, also known as material objectivity is one of the most discussed fundamental principles of classical continuum physics. It appears evident that material properties are independent of frames of reference. The principle has several different mathematical realizations. 

The formulation of Noll is based on the transformation properties of vectors in a three-dimensional Euclidean vector space, a mathematical representation of the spatial part of our spacetime \cite{TruNol65b,Nol67a,Nol04m}. This formulation has been criticized and reformulated by others \cite{MusRes08a,Fre09a}. The correct method is fundamental, as it is demonstrated by its role in the well-posedness of partial differential equations of Rational Extended Thermodynamics \cite{MulRug98b} or in the  case of extended heat conduction in moving media \cite{Chr09a,Ang23a}. A geometric formulation requires a reference frame independent representation of the corresponding functions and relations in a flat, four dimensional model of our nonrelativistic spacetime \cite{Mat86a,Nol04m}. Some properties of the two approaches are compared in  \cite{MatVan06a,MatVan07a}. In \cite{VanEta19a} the transformation properties of the moment series expansion of Rational Extended Thermodynamics are explained in terms of transformation properties derived in a four-dimensional Galilean relativistic spacetime model. 

The difference between the two approaches is significant. Noll's formulation is based on rigid rotating reference frames and   requires invariance with respect to transformations between these frames. In contrast, Matolcsi's spacetime approach requires a reference frame independent formulation of material properties and transformations between reference frames are derived, secondary concepts.  

Why is this a difficult problem? Firstly, in continuum theory, there is a distinguished reference body: the continuum itself. In this context, a spacetime-based formulation is far from evident, since the properties of the matter must be formulated in a way that takes into account the existence of the matter itself. This is particularly important in dissipative processes, since energy dissipation is objective, measurable, technologically significant and costly. Secondly, in nonrelativistic physics time passes independently of observers, therefore space and spacelike parts of four tensors appear independent of time. 

The two approaches can be compared by its consequences as well. A good example is the thermodynamic consistency of weakly nonlocal Korteweg-type fluids. Dunn and Serrin, \cite{DunSer85a}, applies Noll's concept of objectivity and combines with the Coleman-Noll procedure to derive the constitutive function of the Korteweg pressure. They obtain
\begin{align}\label{DS}
	\qqP_{\rm Dunn-Serrin} &= \left( \qrho^2 \pder{\psi}{\qrho} - \qrho \nabla \cdot   \ppder{\qrho \psi}{\nabla \qrho}  \right) \ident + \nabla\qrho\circ\ppder{\qrho \psi}{\nabla \qrho}, 
\end{align}
where $\qrho$ is the density, $p \qf$ is the fluid pressure, $\psi$ is called free energy in the paper of Dunn and Serrin and $\nabla$ denotes the spatial derivative.  

In the special case of  $\psi = K \frac{(\nabla \qrho)^2}{2 \qrho}$, with $K= \rm const.$ one gets 
\begin{align}\label{qDS}
\qqP_{\rm  DS-quadratidc} &= \left( \qrho K \Delta \qrho - \frac{K}{2}(\nabla\qrho)^2 \right) \ident + K\nabla \qrho\circ \nabla \qrho, 
\end{align}
which is the pressure for diffuse interfaces derived in the framework of phase field models as well \cite{AndEta98a}	
	
On the other hand a different pressure tensor was derived in \cite{Van23a}:
\begin{align}\label{st}
	\qqP_{\rm  st} &= \left(\qrho^2 \pder{u}{\qrho} -\frac{\qrho^2}{2} \nabla \cdot  \ppder{u}{\nabla \qrho}  \right) \ident - \frac{\qrho^2}{2} \nabla \ppder{u}{\nabla \qrho}, 
\end{align}
where $u$ is the specific internal energy.  

In the special case of  $u = K \frac{(\nabla \qrho)^2}{2 \qrho}$, with $K=\rm const.$ one obtains 
\begin{align}\label{qst}
	\qqP_{\rm  st-quadratic} &= \left( -\frac{\qrho K}{2} \Delta \qrho \right) \ident -\frac{\qrho K}{2}\nabla\circ\nabla \qrho+ \frac{K}{2}\nabla \qrho\circ \nabla \qrho.  
\end{align}

Moreover, in \cite{Van23a} the material frame indifference is hidden, relative quantities are used to characterize the fluid, like current densities and relative velocity. Also \re{DS} and \re{st} are different: in \re{DS} interstitial working is introduced, in \re{st} the entropy flux is a constitutive function\footnote{One can get the same result with constitutive entropy flux as well \cite{Mor23a}}. Both papers use rigorous methods: Liu procedure in \re{st} and Colemann-Noll procedure in \re{DS}.  

In the following we shortly compare the objectivity arguments and analyze the reason of the difference of the above pressure tensors. 
	
\section{Galilean relativity}

Noll's concept of objectivity is based on specific transformation rules between reference frames. Covariance with respect to these transformations is a necessary condition for material frame indifference. However, this is not sufficient: reference frame independence requires covariance with respect to all transformations. Nevertheless,  it is difficult to find  experimental counterexamples,  so the principle appears to hold true.  A reference frame independent spacetime approach, however, is different. In this approach, transformation rules can be derived between any reference frames. Also, a geometrical formulation is not always necessary. One can be sure that the theory is Galilean covariant even though it seems to be composed of relative elements.


Spacetime, including the non-relativistic one, is four-dimensional. There are several known formulations (see e.g. \cite{Hav64a,Fri83b,Fre09a,Mat20b,MatEta24m}), and there are some known consequences, for example the covariance of the Maxwell equations in Galilean electrodynamics \cite{Rou13a}. In non-relativistic continua, the usual framework with relative, reference frame dependent quantities result in good models of observed phenomena. However, the story of the principle of material frame indifference indicates that a clear mathematical formulation without conceptual clarity is not enough. Despite absolute time, four-dimensional spacetime concepts are unavoidable in non-relativistic physics to remove relative quantities and reference frame dependence. Their conscious usage clarifies several concepts, like the conductive and convective current densities in local and substantial balances: their relation is a Galilean transformation between the laboratory and material frames. Rigorous spacetime treatment of nonrelativistic fluid mechanics reveals that the energy-momentum tensor is more sophisticated than in special relativity \cite{Mat86a,Van17a}. There is a covariant concept of energy, but only as part of a third-order four-tensor.

However, keeping in mind the spacetime background, it is easy to avoid problematic aspects and keep the theory covariant despite using relative quantities. For example, the spatial derivatives (\ie the gradients) are legitimate, objective constitutive variables; they are spacelike covectors, that do not transform when the reference frame changes \cite{Mat20b}. Moreover, the balance form is a four-divergence of a four-quantity (scalar, vector or tensor), therefore it is invariant, that does not transform either \cite{Van17a}. In addition, the entropy current density is a four-vector, therefore, if its timelike part, the entropy density, is a constitutive quantity, then its spacelike part, the entropy current density, must also be a constitutive quantity.


Therefore, keeping in mind that the
\begin{itemize}
	\item constitutive state space is spanned by thermodynamic state variables and its gradients,
	\item the constraints are fundamental balances,
	\item entropy flux is a constitutive quantity,
\end{itemize}
one can conclude that Korteweg fluids are safe from the point of view of material frame indifference. The final proof to this statement could be the development of a reference frame independent Galilean covariant Korteweg fluid theory. 

\section{Thermodynamic methodology}

Fluid mechanics starts with the fundamental balances of mass, momentum and energy:
\begin{align}
	\label{eq:bal-m}
	\dot \qrho + \qrho \nabla \cdot \qqv &= 0 , \\
	\label{eq:bal-v}
	\qrho \dot \qqv + \nabla \cdot \qqP &= 0 , \\
	\label{eq:bal-e}
	\qrho \dot e + \nabla \cdot \qqJE &= 0 ,
\end{align}
Eulerian description is applied, dot denotes substantial time derivatives and $ \qrho, \qqv, e $ are the density, velocity and the  (mass)specific energy of the fluid. The specific internal energy is  $ u = e - \frac{1}{2} \qqv \cdot \qqv $. $ \qqP $ is the pressure tensor, which is symmetric as a consequence of conservation of angular momentum and $ \qqJE $ is the conductive current density of the total energy. The entropy balance is
\begin{align}
	\label{eq:s-ineq}
	0 \le \qrho \dot s + \nabla \cdot \qqJS ,
\end{align}
where $ s $ and $ \qqJS $ denote the mass-specific entropy function and the entropy current density, respectively. $ \qqP, \qqJE, \qqJS $ and $s$ are constitutive quantities, depend on the following weakly nonlocal constitutive state space:

\begin{align}
	\label{eq:const-st-sp}
	\Big( \qrho , \nabla \qrho , \nabla \otimes \nabla \qrho , \qqv , \nabla \qqv , e , \nabla e  \Big) .
\end{align}

Introducing the  $\br,  \bv, \be $ Lagrange-Farkas multipliers for the balances above and $\bgr{}$ for the spatial derivative of \re{eq:bal-m} one can apply Liu procedure to obtain the consequences of the entropy inequality. The derivative of the mass balance is a constraint, because the constitutive state space is second order weakly nonlocal in the density \cite{Van05a,Cim07a}. Then Liu procedure leads to the following inequality (23)\footnote{Abstract index notation is introduced for the three vectors and tensors. Double indexes denote contraction. }:		
\begin{align}
			\nonumber
			0 &\le \left( \qrho \pder{s}{\qrho} - \br \right) \underline{\dot \qrho} + 
			\left( \qrho \ppder{s}{\pd_i \qrho} - \bgr{i} \right) \underline{\qdot{\pd_i \qrho}} + 
			\qrho \ppder{s}{\pd_{ij} \qrho} \underline{\qdot{\pd_{ij} \qrho}}  
			\\
			\nonumber
			& \hskip 2.7ex + \left( \pder{s}{\qv^i} - \bv{i} \right) \qrho \underline{\dot \qv^i} + \qrho \ppder{s}{\pd_j \qv^i}\underline{\qdot{\pd_j \qv^i}} + \left( \pder{s}{e} - \be \right) \qrho \underline{\dot e} + \qrho \ppder{s}{\pd_i e} \underline{\qdot{\pd_i e}}  \\
			\nonumber
			& \hskip 2.7ex + \left( \pder{\qJS^j}{\qrho} - \bv{i} \pder{\qP^{ij}}{\qrho} - \be \pder{\qJE^j}{\qrho}  \right) \pd_j \qrho + \left( \ppder{\qJS^j}{\pd_k \qrho} - \bv{i} \ppder{\qP^{ij}}{\pd_k \qrho} - \be \ppder{\qJE^j}{\pd_k \qrho}  \right) \pd_{jk} \qrho \\
			\nonumber
			& \hskip 2.7ex + \left( \ppder{\qJS^j}{\pd_{kl} \qrho} - \bv{i} \ppder{\qP^{ij}}{\pd_{kl} \qrho} - \be \ppder{\qJE^j}{\pd_{kl} \qrho}  \right) \underline{\pd_{jkl} \qrho} + \left( \pder{\qJS^j}{\qv^i} - \bv{l} \pder{\qP^{lj}}{\qv^i} - \be \pder{\qJE^j}{\qv^i} \right) \pd_j \qv^i \\
			\nonumber
			& \hskip 2.7ex + \left( \ppder{\qJS^j}{\pd_k \qv^i} - \boxed{\frac{\qrho}{2} \bgr{l} \left( \delta_i^j \delta_l^k + \delta_i^k \delta_l^j \right)} - \bv{l} \ppder{\qP^{lj}}{\pd_k \qv^i} - \be \ppder{\qJE^j}{\pd_k \qv^i}  \right) \underline{\pd_{jk} \qv^i} \\
			\nonumber
			& \hskip 2.7ex + \left( \pder{\qJS^j}{e} - \bv{i} \pder{\qP^{ij}}{e} - \be \pder{\qJE^j}{e}  \right) \pd_j e + \left( \ppder{\qJS^j}{\pd_k e} - \bv{i} \ppder{\qP^{ij}}{\pd_k e} - \be \ppder{\qJE^j}{\pd_k e}  \right) \underline{\pd_{jk} e} \\
			\label{eq-s-ineq-nonloc}
				& \hskip 2.7 ex - \left( \qrho \br \delta_j^i + \bgr{i} \pd_j \qrho  \right) \pd_i \qv^j - \bgr{i} \pd_i \qrho \pd_j \qv^j .
		\end{align}

		Here the boxed expression takes the symmetric part of $\qrho \bgr{k} \delta_i^j$, as it is required by the symmetry of the second derivative of the velocity, $\pd_{jk} \qv^i$ \cite{Cim02a}. The Liu equations are the coefficients of the underlined terms. 
		Solving them and introducing the extra energy $\varepsilon$:
		\begin{align}
			\label{eq:e-nonloc}
			u \qf = e-\frac{v^2}{2} = 
			u  -  \frac{\he \left( \qrho , \nabla \qrho  \right) }{\qrho},
		\end{align}
		 leads to the following form of the residual inequality \cite{Van23a}:
		\begin{align}
			\nonumber
			0 &\le
		\left( \qqJE - \qqP \cdot \qqv - \frac{\qrho}{2} \left( \ppder{\he}{\nabla \qrho} \left( \nabla \cdot \qqv \right) + \ppder{\he}{\nabla \qrho} \cdot \nabla \qqv \right) \right) \cdot \nabla \frac{1}{T}  \\
			\label{eq:dis-ineq-nonloc-2}
			& \hskip 2.7ex
			- \frac{1}{T} \left( \qqP - \left[\left( p \qf - \he + \qrho \pder{\he}{\qrho} - \frac{\qrho^2}{2} \nabla \cdot \left( \frac{1}{\qrho} \ppder{\he}{\nabla \qrho} \right) \right) \ident 
			- \frac{\qrho^2}{2} \nabla \left( \frac{1}{\qrho} \ppder{\he}{\nabla \qrho} \right)\right]  \right) : \nabla \qqv .
		\end{align}
		
		If the necessity of symmetrization is not realized, leaving the related term as it appears after the composite derivation we obtain for the boxed part the following expression:
		$$
			\boxed{\qrho \bgr{l} \delta_i^j \delta_l^k}\pd_{jk} \qv^i =\qrho \bgr{k}\pd_{ik} \qv^i.	
		$$
	The residual inequality will be different than above:		
	\begin{align}
		\nonumber
		0 &\le
		\left( \qqJE - \qqP \cdot \qqv - \qrho  \ppder{\he}{\nabla \qrho}  \nabla \cdot \qqv \right) \cdot \nabla \frac{1}{T}  \\
		\label{eq:dis-ineq-nonloc-1}
		& \hskip 2.7ex
		- \frac{1}{T} \left( \qqP - \left[\left( p \qf - \he + \qrho \pder{\he}{\qrho} - \qrho \nabla \cdot   \ppder{\he}{\nabla \qrho}  \right) \ident - \nabla\qrho\circ\ppder{\he}{\nabla \qrho}\right]   \right) : \nabla \qqv .
	\end{align}
	
	In the symmetrized and non symmetrized cases the pressure tensors of a perfect fluid are respectively:
		\begin{align}
	  \nonumber
	\qqP_{\rm symm} &= \left( p \qf - \he + \qrho \pder{\he}{\qrho} - \frac{\qrho^2}{2} \nabla \cdot \left( \frac{1}{\qrho} \ppder{\he}{\nabla \qrho} \right) \right) \ident + \frac{\qrho^2}{2} \nabla \left( \frac{1}{\qrho} \ppder{\he}{\nabla \qrho} \right)\\	\nonumber		
	 \qqP_{\rm asymm} &= \left( p \qf - \he + \qrho \pder{\he}{\qrho} - \qrho \nabla \cdot   \ppder{\he}{\nabla \qrho}  \right) \ident - \nabla\qrho\circ\ppder{\he}{\nabla \qrho} 
	 	\end{align}
	
	The divergence of the two tensors are the same,
	\begin{align}
		\label{eq:holop}
		\nabla \cdot \qqP \id = \nabla p \qf + \qrho \nabla \frac{\delta \he}{\delta \qrho}.
	\end{align}
	One can see that both are classically holographic with the terminology of \cite{Van23a} if one applies the Gibbs-Duhem relation for the gradient of the static pressure. Introducing the specific extra energy $\psi = \he/\qrho$ one obtains the Dunn-Serrin form, \re{DS} and also \cite{DunSer85a} (5.3), after the transformation from $\qqP_{\rm asymm}$. The pressure tensor \re{st} and also \cite{Van23a} (65) comes from $\qqP_{\rm symm}$. 
	
\section{Discussion}	

\subsection{Objectivity} Dunn and Serrin's work is not restricted to fluids; they begin their analysis within the framework of finite deformation elasticity. Several examples demonstrate that, in the case of fluids, explicit objectivity checking can be avoided and relative quantities can be used to obtain the same Korteweg stress. We have argued here that a background geometric Galilean relativistic flat spacetime model provides an explanation: the standard fluid mechanics practices, supplemented by the conscious use of the basic properties of the spacetime model (e.g. that gradients are frame independent), offer a safe haven for objective calculations. A genuinely Galilean relativistic spacetime-based Korteweg fluid theory  is lacking yet.

\subsection{Various Korteweg pressures} The difference in the ideal Korteweg fluid pressures  in references \cite{DunSer85a} and \cite{Van23a} is due to a symmetry requirement in the mathematical formulation of the entropy principle. The requirement remains hidden in the variational approaches and in case of the Coleman-Noll procedure, but becomes apparent when the Liu procedure is applied. The application of the symmetry requirement in \cite{Van23a} is not complete either (see \cite{Cim02a}).

\subsection{Diffuse interfaces and microforce balance} The Korteweg pressure forms the theoretical basis of diffuse interfaces, arising as model H in phase field theory \cite{AndEta98a,HohHal77a}. There it is derived from the Noether theorem using a variational principle. The concept of microforce balance was introduced as an alternative to the variational principles of phase field theories \cite{Gur96a}. However, both concepts are superfluous in the light of a rigorous  thermodynamic approach, particularly with regard to the classical holographic property, which is a consequence of the second law \cite{SzuVan25m}. It is remarkable, that both symmetric and assymmetric forms of the pressure can be compatible by the Enskog-Vlasov equation, due to the classical holographic property of the Vlasov force \cite{StrFre22a} equation (4.5).

\section{Acknowledgement}   
		
The author thank Henning Struchtrup for recognising the difference between the pressure tensors and thank the Galileo Galilei Institute for Theoretical Physics for the hospitality and the INFN for partial support during the preparation of this work. 

The author thank NKFIH NKKP-Advanced 150038 for support. 
		
The research reported in this paper is part of project no. TKP-6-6/PALY-2021, implemented with the support provided by the Ministry of Culture and Innovation of Hungary from the National Research, Development and Innovation Fund, financed under the TKP2021-NVA funding scheme.
		
\bibliographystyle{unsrt}

\end{document}